\documentclass[prl, twocolumn, showpacs]{revtex4-1}
\usepackage{float}
\usepackage{bbm}
\usepackage{epsfig}
\usepackage{epstopdf}
\usepackage{graphicx}
\usepackage{amsmath,amssymb}
\usepackage{color}
\usepackage{hyperref}
\usepackage{caption}
\usepackage{subcaption}
\usepackage{ragged2e}
\usepackage{bibunits}
\begin{document}

\begin{bibunit}[apsrev4-1]

\setlength{\abovedisplayskip}{5pt}
\setlength{\belowdisplayskip}{5pt}

\title{Dynamically Correcting a {\sc cnot} Gate for any Systematic Logical Error}

\author{F.~A.~Calderon-Vargas}
\email{f.calderon@umbc.edu}
\author{J.~P.~Kestner}
\affiliation{Department of Physics, University of Maryland Baltimore County, Baltimore, Maryland 21250, USA}

\begin{abstract}
We derive a set of composite pulse sequences that generates {\sc cnot} gates and correct all systematic errors within the logical subspace to arbitrary order. These sequences are applicable for any two-qubit interaction Hamiltonian, and make no assumptions about the underlying noise mechanism except that it is constant on the time scale of the operation.  We do assume access to high-fidelity single-qubit gates, so single-qubit gate imperfections eventually limit the achievable fidelity.  However, since single-qubit gates generally have much higher fidelities than two-qubit gates in practice, these pulse sequences offer useful dynamical correction for a wide range of coupled qubit systems.
\end{abstract}

\maketitle
\begin{figure}
  \centering
   \captionsetup{font=small,justification=RaggedRight}
  \includegraphics[trim=0cm 3.7cm 0cm 3.7cm, clip=true,width=8.5cm, angle=0]{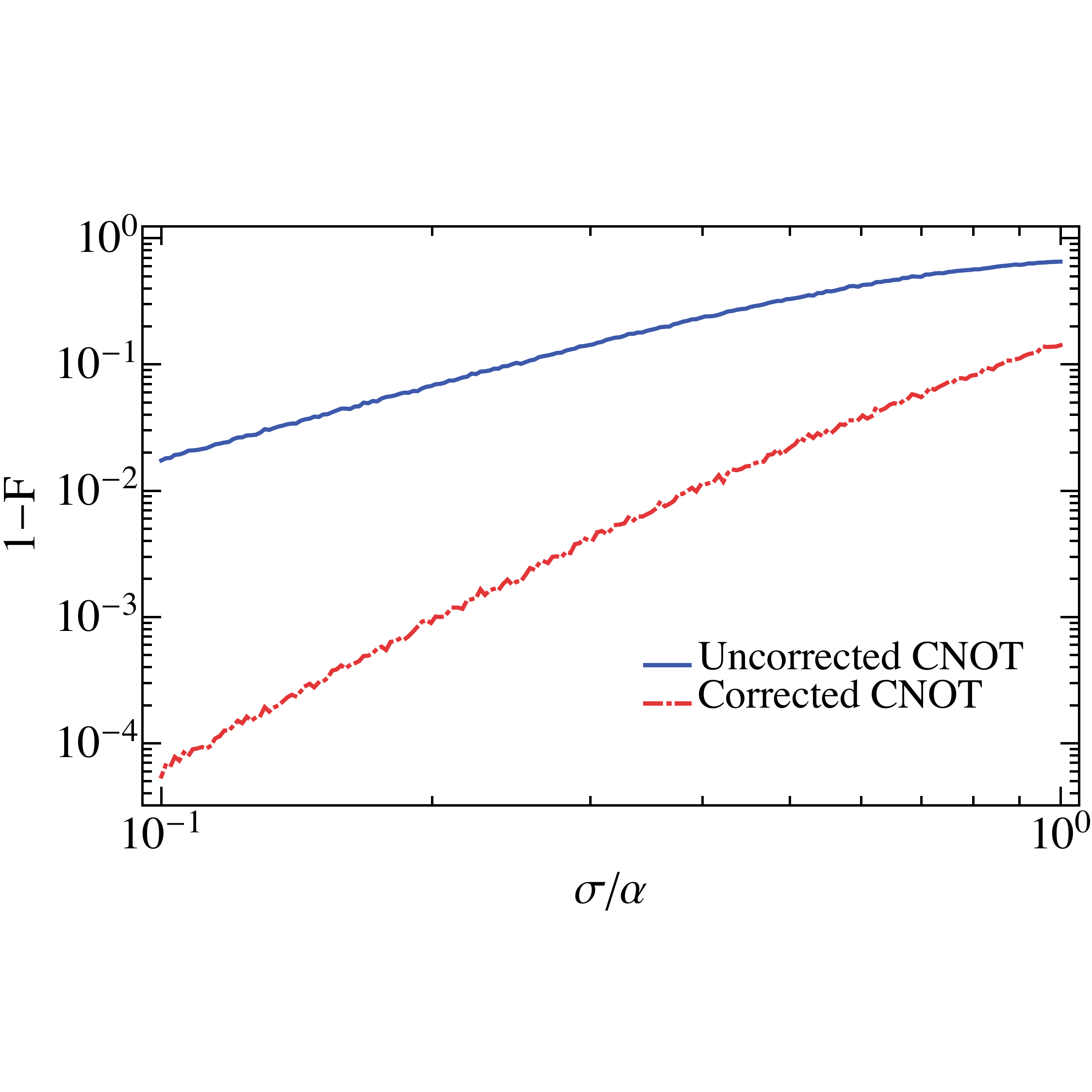}
  \caption{Infidelity vs noise strength for uncorrected (solid curve) and corrected (dashed curve) {\sc cnot} gates for Heisenberg-coupled spins in a random magnetic field.}\label{fig:length5_Infidelity}
\end{figure}
\begin{figure}[tbp]
   \centering
   \captionsetup{font=small,justification=RaggedRight}
   \includegraphics[trim=0cm 3.7cm 0cm 3.7cm, clip=true,width=8.5cm, angle=0]
   {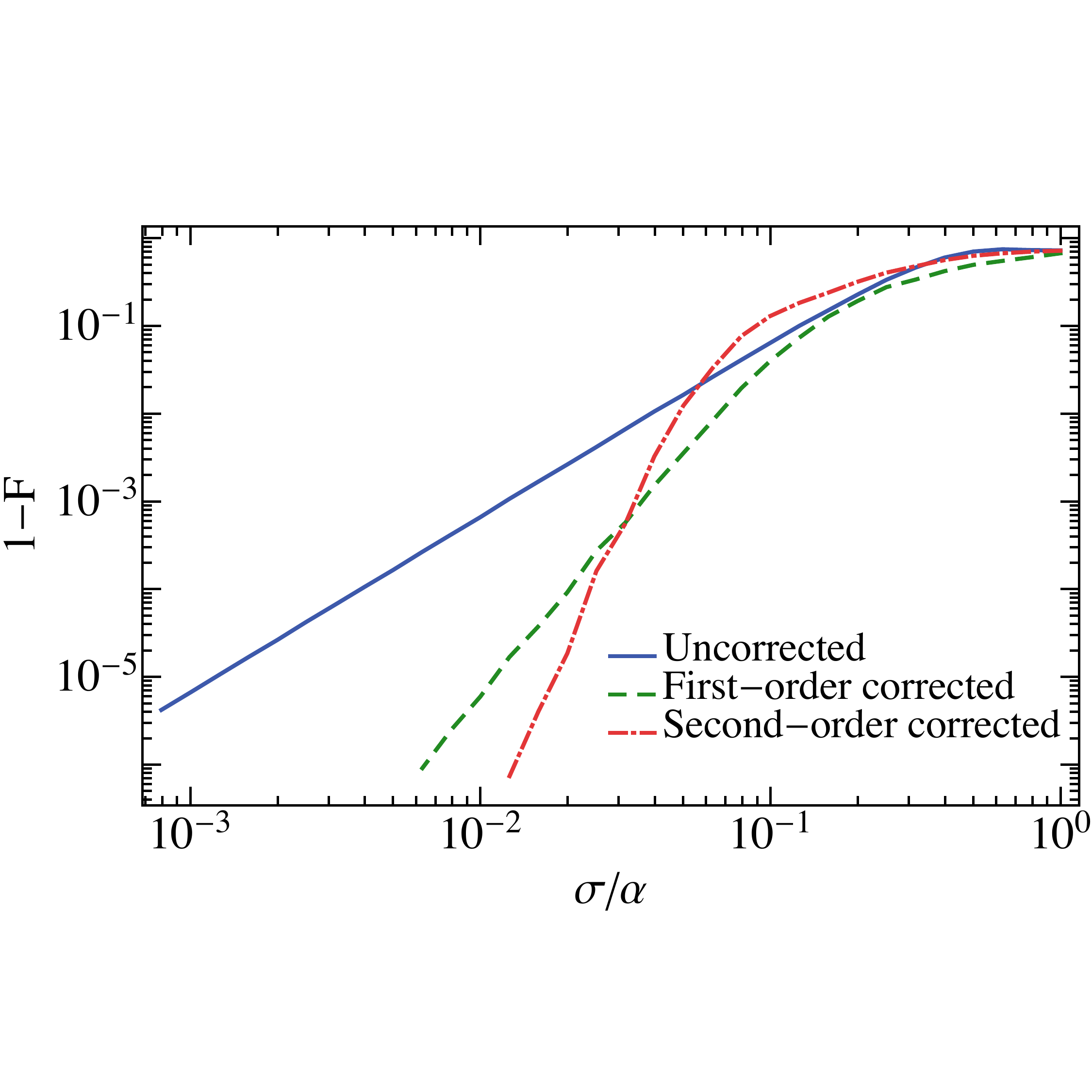} 
   \caption{Infidelity vs noise strength for uncorrected
   (solid curve) and corrected {\sc cnot} gates to first (dashed curve) and second
   (dot-dashed curve) order, using an Ising Hamiltonian with general SU(4) errors.}
   \label{fig: CNOT_20_Piece_Infidelity}
\end{figure}

Scalable fault-tolerant quantum computing requires gate operations with errors below the quantum error correction threshold, with lower errors allowing more efficient scaling.  A significant source of error in many settings is coherent, systematic gate error.  In particular, maximally entangling two-qubit operations tend to have significantly worse fidelities than single-qubit operations, for instance, due to weak interactions requiring a long gate time in the presence of low-frequency noise.  Such unknown systematic errors may be compensated using composite pulse sequences \cite{Levitt1986,Jones2011,Merrill2012,Lidar2013}, where a desired operation is replaced by a set of imperfect pulses in such a way that the systematic errors inherent in each pulse cancel each other. A feature of composite sequences is that their analytical construction allows great generality, and they have been applied in NMR \cite{Vandersypen2005}, trapped ions \cite{Mount2015}, nitrogen-vacancy centers in diamond \cite{Ryan2010}, electron spins in semiconductor quantum dots \cite{Eng2015}, quantum optics \cite{Genov2014}, atom interferometry \cite{Berg2015}, etc.
\\
\indent However, most of the progress with this approach has been made on single-qubit gates, and there is no general two-qubit sequence able to produce an entangling gate corrected to leading order in every possible systematic error.  In fact, there is a ``no-go" theorem for black-box dynamical correction \cite{Khodjasteh2009b}, requiring detailed knowledge of the relationships between the errors of each component of a composite sequence.  Some sequences circumvent this by correcting only a limited subset of errors using an isomorphism between the SU(2) generators and a subgroup of the SU(4) generators \cite{Jones2003,Tomita2010}. Alternatively, the term isolation approach \cite{Hill2007} recovers the generality of the error suppression, but by a factor of the inverse of the number of pulses rather than nulling first- or higher-order terms in the noise, and that requires many more pulses to achieve the same level of performance as an order-by-order approach at small error rates.
\\
\indent Another approach is to use optimal control theory to numerically optimize performance for a specific system \cite{Palao2002,Khaneja2005,Machnes2011}. This has been highly successful, particularly in NMR \cite{Lu2015}, and the resulting pulses, though they rarely provide much physical insight, tend to be much shorter, lower power, and produce higher gate fidelities at the cost of system-independence. One must be able to write down the full master equation in order to do the numerics, and unknown correlations between the noise and the control cannot be handled in an open-loop process.  Such numerical techniques also can be complicated by control constraints that introduce local traps into the search landscape \cite{Riviello2015}. We shall focus on the complementary analytical approach.  Thus, the prime use for our results will be in cases where one has limited knowledge of the underlying physical noise and control mechanisms, highly constrained control, or both, especially in an open-loop setting.
\\
\indent In this Letter, we present a family of general composite pulse sequences that generate {\sc cnot} gates compensated for all systematic errors to arbitrary order using as a building block any single imperfect entangling gate, while only assuming access to high-fidelity single-qubit operations.  The assumption of negligible errors in the local gates opens a loophole in the no-go theorem \cite{Khodjasteh2009b}, similar in spirit to the exploitation of a ``robust operating point" in Ref.~\cite{Witzel2015}.  The purpose of employing only a single nonlocal gate is to guarantee that all systematic errors entering the sequence are identical, though unknown.  This makes our treatment highly compatible and modular, being impervious to the details of the qubit system and the method by which the gates are performed. Thus, unlike the application of standard local dynamical decoupling sequences during entanglement \cite{DArrigo2016}, our approach does not require local noise nor instantaneous pulses.  Each part of the sequence may be performed in whatever way is experimentally convenient; the modular construction does not assume any particular form. This feature allows it to be used in conjunction with optimal control theory or other dynamical correction methods if desired, for instance, using numerically shaped pulses to optimally produce the components of the sequence or else using our sequence to inform the trial pulse input to a numerical search algorithm, or even provide a second boost to the two-qubit gate fidelity after an initial  improvement using particular experimental techniques such as active cancellation tones \cite{Sheldon2016}.
\\
\indent The pulse sequences are composed using repetitions of the nonlocal gate $\left(\theta\right)_{ZZ}=\exp\left[-i\frac{\theta}{2} \sigma_{ZZ}\right]$, which, regardless of the form of the two-qubit interaction Hamiltonian, can be generated by one or more applications of the evolution operator \cite{Zhang2003,Geller2010} along with appropriate single-qubit gates obtained from Cartan decomposition \cite{Kraus2001}.  The family of composite pulse sequences is formed by the sequential application of $\left(\theta\right)_{ZZ}$ interleaved with local $\pi$ rotations $\sigma_{ij}\equiv \sigma_i \otimes\sigma_j$, with $i,j=\{I,X,Y,Z\}$. In this Letter, we first introduce various composite sequences capable of correcting various subsets of systematic errors, and then we nest those sequences to form a completely general sequence that corrects any systematic error and generates a first-order error-free {\sc cnot} gate using no more than 120 nonlocal gates. This sequence can itself be nested to correct errors up to any order.
\\
\indent As a starting point, we represent the error in a noisy realization of the building block $\left(\theta\right)_{ZZ}$ by expanding it to first order as
\begin{equation}\label{eq: error_expansion}
\left(\theta\right)_{ZZ}=\exp\left[-i\frac{\theta}{2} \sigma_{ZZ}\right]\left(I+i\underset{i,j=\{I,X,Y,Z\} }{\sum}\delta_{ij}\sigma_{ij}\right),
\end{equation}
where $\delta_{ij}$ is hereafter referred to as the error term in the \textit{ij} error channel.
\\
\indent As a warm-up, consider the simplest possible sequence, formed by inserting a local $\pi$ rotation, $\sigma_{ab}$, hereafter referred to as an echo pulse, in between two applications of the noisy entangling operation, $\left(\theta\right)_{ZZ}$.  To generate an entangling operation, one should choose a $\sigma_{ab}$ that commutes with $\sigma_{ZZ}$ since the anticommuting alternative would simply produce a purely local operation.  The result, up to first order in the errors, is
\begin{multline}\label{eq: two-piece_sequence}
\mathcal{U}_{ab}^{\left(2\right)}\left[\left(\theta\right)_{ZZ}\right]=\left(\theta\right)_{ZZ}\sigma_{ab}\left(\theta\right)_{ZZ}\sigma_{ab}=\exp\left[-i\frac{2\theta}{2} \sigma_{ZZ}\right] \\
\times\left\{I+i\underset{i,j}{\sum}\delta_{ij}\left(\sigma_{ab}\sigma_{ij}\sigma_{ab}+\left(\theta\right)_{ZZ}^{\dagger}\sigma_{ij}\left(\theta\right)_{ZZ}\right)\right\},
\end{multline}
where the bracketed term on the lhs indicates the nonlocal rotation used to build the sequence. Hence, this sequence corrects all error channels that simultaneously commute with the entangling operation and anticommute with the echo pulse, $\left[\sigma_{ij},\sigma_{ZZ}\right]=\left\{\sigma_{ij},\sigma_{ab}\right\}=0$.  In fact, this is clearly true not just to first order but to all orders.  This simple sequence is already known \cite{Viola1999} and has appeared in both theory \cite{Hill2007} and experimental work \cite{Shulman2012,Chow2013}. We call this a length-2 sequence, where by ``length-$n$" we mean a sequence having $n$ applications of the noisy entangling operation.
\\
\indent Furthermore, by placing this length-2 pulse sequence \eqref{eq: two-piece_sequence} inside a second length-2 sequence that uses an echo pulse that anticommutes with the first pulse, one produces a length-4 sequence that exactly cancels all error channels that commute with $\sigma_{ZZ}$, excluding the coupling error $\delta^{ZZ}$ itself.   (Note that it is thus not actually necessary for the controlled part of the building block to be strictly a $ZZ$ rotation when using the length-4 sequence, as long as the other generators in the exponent commute with $\sigma_{ZZ}$.)  Choosing $\theta=\pi/4$ for length-2 or $\theta=\pi/8$ for length-4, the result of the sequence is locally equivalent to a {\sc cnot}. Then it is straightforward to apply the two-qubit variant of the BB1 pulse sequence \cite{Jones2003} to also cancel the $ZZ$ error channel to second order.\\
\indent Now we turn our attention to canceling, to leading order, the error channels that do not commute with $\sigma_{ZZ}$. By analogy with the previous case, we focus our attention on a sequence of the form
\begin{multline}\label{eq: n-piece_sequence}
\sigma^{\left(n\right)}_{\mathrm{echo}}\left(\theta\right)_{ZZ}\sigma^{\left(n\right)}_{\mathrm{echo}}\sigma^{\left(n-1\right)}_{\mathrm{echo}}\left(\theta\right)_{ZZ}\sigma^{\left(n-1\right)}_{\mathrm{echo}}\ldots\sigma^{\left(1\right)}_{\mathrm{echo}}\left(\theta\right)_{ZZ}\sigma^{\left(1\right)}_{\mathrm{echo}}
\\
=\exp\left[-i\frac{\theta}{2}\left(\overset{n}{\underset{l=1}{\sum}}\xi_l\right)\sigma_{ZZ}\right]
\\
\times\left\lbrace I+i\underset{i,j}{\sum}'\delta_{ij}\sigma_{ij}\overset{n}{\underset{m=1}{\sum}}\zeta_m^{ij}\exp\left[-i\theta\left(\overset{m-1}{\underset{l=1}{\sum}} \xi_l \right) \sigma_{ZZ}\right] \right\rbrace,
\end{multline}
where $\sigma^{\left(m\right)}_{\mathrm{echo}}$ denotes local $\pi$ rotations of the form $\sigma_{ab}$, the primed sum indicates that we include only error channels that anticommute with $\sigma_{ZZ}$, and
\begin{equation}\label{eq: xi}
 \xi_l \equiv
  \begin{cases}
   +1 & \text{if } \left[\sigma^{\left(l\right)}_{\mathrm{echo}},\sigma_{ZZ}\right] = 0 \\
   -1 & \text{if } \lbrace\sigma^{\left(l\right)}_{\mathrm{echo}},\sigma_{ZZ}\rbrace = 0
  \end{cases},
\end{equation}
\begin{equation}\label{eq: zeta}
 \zeta_m^{ij} \equiv
  \begin{cases}
   +1 & \text{if } \left[\sigma^{\left(m\right)}_{\mathrm{echo}},\sigma_{ij}\right] = 0 \\
   -1 & \text{if } \lbrace\sigma^{\left(m\right)}_{\mathrm{echo}},\sigma_{ij}\rbrace = 0.
  \end{cases}
\end{equation}
\indent Setting the real and imaginary parts of the error term in Eq. \eqref{eq: n-piece_sequence} to equal zero requires the two equations
\begin{equation}\label{eq: set_of_eqs_error_n_piece}
\begin{aligned}
\zeta_1^{ij}+\overset{n}{\underset{m=2}{\sum}}\zeta_m^{ij}\cos\left[\overset{m-1}{\underset{l=1}{\sum}} \xi_l \theta \right] &= 0\\
\overset{n}{\underset{m=2}{\sum}}\zeta_m^{ij}\sin\left[\overset{m-1}{\underset{l=1}{\sum}} \xi_l \theta \right] &= 0
\end{aligned}
\end{equation}
to hold for each error channel corrected.
For lengths $n=3$ and $n=4$, we have found solutions that cancel some but not all error terms. With $n=5$, though, by choosing echo pulses such that all $\xi_l=1$, $\zeta_2^{ij}=\zeta_4^{ij}$, $\zeta_1^{ij}=\zeta_5^{ij}$, and using Chebyshev's recursive formula for cosine and sine of multiple angles, we simplify Eq.~\eqref{eq: set_of_eqs_error_n_piece} to the single condition
\begin{equation}\label{eq: 5_piece_error_eq}
\zeta_3^{ij}+\zeta_4^{ij}2\cos\theta+\zeta_5^{ij}\left(4\cos^2\theta-2\right)=0.
\end{equation}
There are real solutions for $\theta$ that satisfy this equation for all error terms as long as the echo pulses are taken such that either $\zeta_3^{ij}=\mp1$ and $\zeta_4^{ij}=\zeta_5^{ij}=\pm 1$ or $\zeta_5^{ij}=\mp1$ and $\zeta_3^{ij}=\zeta_4^{ij}=\pm 1$; we proceed with the former choice since it gives the smaller value of $\theta$ and hence, presumably, the faster implementation.  This solution is $\theta_0=\arccos\left[\frac{1}{4}\left(\sqrt{13}-1\right)\right] \approx 0.27\pi$. The corresponding $\sigma_{\text{echo}}^{\left(l\right)}$ in Eq. \eqref{eq: n-piece_sequence} are $\sigma_{\text{echo}}^{\left(1,2,4,5\right)}=I$ and $\sigma_{\text{echo}}^{\left(3\right)}=\sigma_{ZZ}$. Therefore, we have found a length-5 sequence that corrects all anticommuting error channels to first order:
\begin{equation}\label{eq: 5-piece_sequence}
\begin{aligned}
\mathcal{U}^{\left(5\right)}\left[\left(\theta_0\right)_{ZZ}\right]&=\left(\theta_0\right)_{ZZ}\left(\theta_0\right)_{ZZ}\sigma_{ZZ}\left(\theta_0\right)_{ZZ}\sigma_{ZZ}\left(\theta_0\right)_{ZZ}\left(\theta_0\right)_{ZZ}\\
&=\exp\left[-i\frac{5\theta_0}{2}\sigma_{ZZ}\right]\Big (I+\mathcal{O}\left(\delta_{\text{anticomm}}^2\right)\Big ),
\end{aligned}
\end{equation}
where the above neglects commuting errors.\\
\indent Now we combine the length-5 sequence \eqref{eq: 5-piece_sequence} above with the length-2 sequence \eqref{eq: two-piece_sequence} that addresses commuting errors to correct both types of errors at the same time. For instance, nesting a single length-2 sequence, with a total rotation angle equal to $\theta_0 $, within the length-5 pulse sequence produces a length-10 sequence, $\mathcal{U}^{\left(10\right)}\left[\left(\theta_0/2\right)_{ZZ}\right]=\mathcal{U}^{\left(5\right)}\left[\mathcal{U}^{\left(2\right)}_{ab}\left[\left(\theta_0/2\right)_{ZZ}\right]\right]$, with three remaining error terms to leading order: $\delta_{ZZ}$, $\delta_{ab}$, and $\delta_{ZZ \cdot ab}$. Of course, for a specific physical system, if one can arrange for those terms to be negligible, one need not go further.  But one may remove all non-$ZZ$ error channels with a length-20 sequence:
\begin{equation}\label{eq: 20_piece_seq}
\begin{aligned}
\mathcal{U}^{\left(20\right)}\left[\left(\frac{\theta_0}{4}\right)_{ZZ}\right]=&\mathcal{U}^{\left(5\right)}\left[\mathcal{U}^{\left(2\right)}_{XX}\left[\mathcal{U}^{\left(2\right)}_{ZI}\left[\left(\frac{\theta_0}{4}\right)_{ZZ}\right]\right]\right]
\\
=&\exp\left[-i\frac{5\theta_0}{2}\sigma_{ZZ}\right]\Big (I+\mathcal{O}\left(\delta_{\text{non-ZZ}}^2\right)\Big ).
\end{aligned}
\end{equation}
\indent It is worth noting that the order in which we nest the length-2 and length-5 sequences can be interchanged in the combined sequences discussed above, giving us correcting sequences that require a smaller number of local gates (a difference of four and eight local operations, for the length-10 and length-20 sequences, respectively), but with the caveat that the error correction performance suffers due to the asymmetry that the length-2 sequence cancels errors to all orders while the length-5 only to first order, and it is clearly better to repeatedly invoke the more accurate sequence.
\\
\indent While the nonlocal $5\theta_0$ rotation is not equivalent to a {\sc cnot}, two applications of this gate with appropriate single-qubit operations can generate a {\sc cnot} gate:
\begin{equation}\label{eq: pulse_seq_into_CNOT}
\begin{aligned}
\text{{\sc cnot}}=&A_1\exp\left[-i\frac{\psi}{2}\sigma_{XI}\right]\mathcal{U}^{\left(k\right)}\exp\left[-i\frac{\phi}{2}\sigma_{XI}\right]\\
&\times \mathcal{U}^{\left(k\right)}\exp\left[-i\frac{\psi}{2}\sigma_{XI}\right]A_2,
\end{aligned}
\end{equation}
where $k=5$, 10, or 20 and the local gates are given by $A_1=\exp\left[\frac{-i\pi\left(\sigma_{X}-\sigma_{Y}\right)}{2\sqrt{2}} \right] \otimes \exp\left[ \frac{-i5\pi\left(\sigma_{X}+\sigma_{Y}-\sigma_{Z}\right)}{3\sqrt{3}}\right]$, $A_2=\sigma_{X} \otimes \exp\left[ \frac{-i\pi\sigma_{Y}}{4}\right]$, $\psi =2\arctan\left[\frac{\sqrt{-57+16\sqrt{13}}}{4-\sqrt{13}+2\sqrt{-7+2\sqrt{13}}}\right]$, and $\phi =-2\arccos\left[-\frac{1}{2\sqrt{-14+4\sqrt{13}}}\right]$.
\\
\indent As a simple example, we consider qubits with an $XYZ$ coupling perturbed by random local terms, $H=\alpha \left(\sigma_{ZZ}+\Delta_X \sigma_{XX}+\Delta_Y \sigma_{YY}\right)+\sum_{j=\{X,Y,Z\}} \left(B_{j,1}\sigma_{jI}+B_{j,2}\sigma_{Ij}\right)$.  This model encompasses both Ising and isotropic Heisenberg couplings, both of which are realized in a broad variety of physical settings.  For instance, one concrete realization would be electron spins in GaAs lateral quantum dots coupled via Heisenberg exchange in the presence of noisy magnetic fields due to nuclear spin fluctuations and the motion of the spin in an applied magnetic field gradient \cite{Brunner2011}.  Figure \ref{fig:length5_Infidelity} shows the average infidelity, defined as in Ref.~\cite{Cabrera2007}, of the uncorrected {\sc cnot} formed in the standard way via two $\sqrt{\text{\sc swap}}$ gates along with local operations \cite{Loss1998}.  The infidelity is averaged over the noise by independently sampling the six random variables over a normal distribution of standard deviation $\sigma$, with the average being taken over 2000 samples for each value of $\sigma$.  The average infidelity of the corrected sequence of Eq.~\eqref{eq: pulse_seq_into_CNOT} with $k=20$ is also plotted, showing impressive reductions in the error with relatively low overhead.  The sequence has 40 coupling pulses for a total interaction time of $5\theta_0/\alpha$, only about three times longer than the total naive interaction time of $\pi/2\alpha$. In Supplemental Material, we present one more example of gate fidelity improvement by applying our length-5 sequence, Eq.~\eqref{eq: pulse_seq_into_CNOT} with $k=5$, to the cross resonance gate between transmon qubits \cite{Sheldon2016}.
\\
\indent Returning to our program, we still have to deal with $\delta_{ZZ}$.  BB1 \cite{Wimperis1994,Jones2003}is a well-known remedy for such errors, but it cannot be applied as the last step here because, after constructing a {\sc cnot} with $\mathcal{U}^{\left(k\right)}$ \eqref{eq: pulse_seq_into_CNOT}, $\delta_{ZZ}$ is no longer just the coefficient of the $ZZ$ error channel but appears in several error channels. Thus one must remove the $\delta_{ZZ}$ term in $\mathcal{U}^{\left(k\right)}$ before it gets mixed into other channels.  Furthermore, BB1 cannot be applied to $\mathcal{U}^{\left(k\right)}$, because BB1 requires the application of nonlocal $\pi$ rotations whose error is proportional to the error in the nonlocal $5\theta_0$ rotation, and this is not possible since $\theta_0$ is not a rational multiple of $\pi$.\\
\indent Nonetheless, the $\delta_{ZZ}$ error in $\mathcal{U}^{\left(k\right)}$ can be compensated using a new composite pulse sequence that is similar in principle to BB1, where we perform the same nonlocal rotation repeatedly about different axes that are tilted from $\sigma_{ZZ}$ toward $\sigma_{ZX}$ by means of inserting single-qubit rotations around $\sigma_{IY}$.  We found numerically that at least five different nonlocal axes are needed to cancel $\delta_{ZZ}$ to first order. The sequence has the form (for $k=5$, 10, or 20)
\begin{equation}\label{eq: ZZ_error_correcting_pulse_seq}
\mathcal{U}^{\left(6k\right)}=\left(\prod_{j=4}^{1}\exp\left[i\frac{\psi_j}{2} \sigma_{IY}\right]\left[\mathcal{U}^{\left(k\right)}\right]^{n_j}\exp\left[-i\frac{\psi_j}{2} \sigma_{IY}\right]\right)\mathcal{U}^{\left(k\right)},
\end{equation}
where the product is in descending order to reflect the time ordering of the operators, $1+\sum n_j=6$, and the values of the parameters depend on what desired operation we are targeting and are obtained from a numerical minimization of an objective function containing the magnitude of the first-order error terms of $\mathcal{U}^{\left(6k\right)}$ as well as the distance between the local invariants \cite{Makhlin2002} of $\mathcal{U}^{\left(6k\right)}$ and those of the target operation.  When targeting a {\sc cnot} equivalent, we obtain a solution $\mathcal{U}^{\left(6k\right)}_{\text{{\sc cnot}}}$ with an intrinsic infidelity (i.e., in the absence of noise) of $\sim 10^{-12}$ given by $n_1=n_2=n_3=1$, $n_4=2$, $\psi_1\approx 1.13527$, $\psi_2\approx -0.40553$, $\psi_3\approx -1.84186$, and $\psi_4\approx 0.19175$.  The local operations that transform that solution into {\sc cnot} are given by
\begin{equation}\label{eq: error-free_CNOT_seq}
\text{{\sc cnot}}= A_1\exp\left[-i\frac{\phi_1}{2} \sigma_{IY}\right]\mathcal{U}^{\left(6k\right)}_{\text{{\sc cnot}}}\exp\left[-i\frac{\phi_2}{2} \sigma_{IY}\right] A_2,
\end{equation}
where $A_1$ and $A_2$ are the same single-qubit operations as in Eq.~\eqref{eq: pulse_seq_into_CNOT}, $\phi_1\approx -1.60782 $, and $\phi_2\approx 0.23403$ (see Supplemental Material).\\
\indent Alternatively, rather than numerically targeting a {\sc cnot}, one can search for parameters such that $\mathcal{U}^{\left(6k\right)}$ instead yields a corrected rotation equivalent to $\left(5\theta_0/k\right)_{ZZ}$.  In that way, the output of the sequence would be a leading-order corrected version of the basic two-qubit input rotation, and thus, in principle, one could correct errors to arbitrarily higher order by nesting the above sequence within itself.  Taking $n_1=n_3=n_4=1$ and $n_2=2$ in Eq.~\eqref{eq: ZZ_error_correcting_pulse_seq}, we do find such numerical solutions with intrinsic infidelities ranging from $10^{-12}$ to $10^{-14}$.  For $k=\{5,10,20\}$, respectively, we find $\psi_1\approx \{-0.18359,-0.10304,-0.05223\}$, $\psi_2\approx \{-3.06178,-3.12993,-3.13844\}$, $\psi_3\approx \{-2.01932,-2.58384,-2.86285\}$, and $\psi_4\approx \{1.75080,0.84439,0.41865\}$.  The necessary local operations to be applied before nesting are given by
\begin{multline}\label{eq: tetha_nested_seq}
\exp\left[-i\frac{5\theta_0/k}{2} \sigma_{ZZ}\right]=\sigma_{XI}^{m_k}\exp\left[-i\frac{\beta_{k}}{2} \sigma_{IY}\right]\\
\times\mathcal{U}^{\left(6k\right)}_{5\theta_0/k}\sigma_{XI}^{m_k}\exp\left[-i\frac{\gamma_{k}}{2} \sigma_{IY}\right],
\end{multline}
where $m_5=m_{10}=1$, $m_{20}=0$, $\beta_{5}\approx 3.11104$, $\gamma_{5}\approx -2.11735$, $\beta_{10}\approx 2.29085$, $\gamma_{10}\approx -1.85051$, $\beta_{20}\approx -1.21618$, and $\gamma_{20}\approx 1.43078$ (see Supplemental Material).\\
\indent We show the efficacy of the full composite pulse sequence by again applying it to a Hamiltonian with an $XYZ$ coupling, but this time with random fluctuations on every one of the 15 SU(4) generators, $H= \alpha\sigma_{ZZ}+\sum \delta_{ij}\sigma_{ij}$. (In the context of our prior example of exchange coupled spins in semiconductor dots, some of the undesired nonlocal terms could, for example, come from a Dzyaloshinskii-Moriya interaction \cite{Dzyaloshinsky1958,Moriya1960}.) Using the length-120 sequence (Eq.~\eqref{eq: error-free_CNOT_seq} with $k=20$), we form a {\sc cnot} that is dynamically corrected against this completely general error to first order. The infidelities are shown for comparison in Fig.~\ref{fig: CNOT_20_Piece_Infidelity}. We also show the infidelity for a second-order corrected {\sc cnot} gate, generated by nesting the sequence from Eq. \eqref{eq: tetha_nested_seq} into Eq. \eqref{eq: error-free_CNOT_seq}.  Again, there are 15 stochastic noise variables, but for the purposes of plotting infidelity we have averaged each one independently over the same normal distribution of standard deviation $\sigma$. In practice, the rather long full composite sequence would be needed only in a worst-case scenario where appreciable systematic error is present in all channels. However, qubits in isotopically enriched silicon (using either gate-defined quantum dots \cite{Veldhorst2015a} or phosphorus donors \cite{Kalra2014}) present long enough coherence times compared to qubit operation times to feasibly implement a sequence of this length.\\
\indent Finally, though we have assumed ideal single-qubit gates throughout (as in Ref.~\cite{Wang2015}), we now discuss the effect imperfect single-qubit gates would have on the above composite sequences. We characterize the effect by assigning to each local gate a random systematic perturbation of the form $\exp\left[-i \sum \Delta_i\sigma_i\right]\otimes \exp\left[-i \sum \Delta_j \sigma_j\right]$ and numerically averaging over noise realizations.
Unsurprisingly, we find that the average infidelity due to imperfect single-qubit gates increases with the sequence length, from three times the average single-qubit infidelity for the length-2 sequence up to 80 times the local gate infidelity for the length-120 sequence, which has 121 local gates (see Supplemental Material for a detailed discussion on the effect of imperfect single-qubit gates with both systematic and random errors). Thus, we expect the various sequences to be useful when the single-qubit gate fidelities are at least an order of magnitude greater than the two-qubit gate fidelities.  However, even in cases where that condition does not hold, the very existence of these sequences reduces the task of raising two-qubit gate fidelity to the considerably simpler task of raising single-qubit gate fidelity.\\
\indent In summary, we have introduced a family of composite pulse sequences capable of correcting any systematic logical error that could appear when generating an entangling gate from any Hamiltonian. We have shown how to use these sequences to generate {\sc cnot} gates that are error-free to arbitrary order. No knowledge of the underlying noise mechanisms is assumed, other than that they are quasistatic on the time scale of the operations.  This sort of black-box approach is generally forbidden \cite{Khodjasteh2009b}, but we have explicitly shown here that it is permitted in the case of access to ideal single-qubit operations.  The generality of the composite pulse sequences we have presented above makes them a powerful tool for robust generation of entanglement and quantum computing in the presence of systematic error.

This work was supported by the National Science Foundation under Grant No.~1620740.  

\putbib[library]

\end{bibunit}

\begin{bibunit}[apsrev4-1]

\onecolumngrid
\clearpage
\begin{center}
\textbf{\large Supplemental Material}
\end{center}
\setcounter{equation}{0}
\setcounter{figure}{0}
\setcounter{table}{0}
\setcounter{page}{1}
\makeatletter
\renewcommand{\theequation}{S\arabic{equation}}
\renewcommand{\thefigure}{S\arabic{figure}}
\renewcommand{\bibnumfmt}[1]{[S#1]}
\renewcommand{\citenumfont}[1]{S#1}

\section{I. \  Objective function used to correct the ZZ-error channel}
Thanks to the isomorphism between SU(2) generators, $\sigma_X,\sigma_Y, \sigma_Z$, and the subgroup of SU(4) generators, $\sigma_{ZZ},\sigma_{ZX},\sigma_{IY}$, the right hand side of Eq. (11) can be expressed in a more tractable way as
\begin{equation}\label{eq:appendix_1}
\left(\prod_{j=4}^{1}\exp\left[i\frac{\psi_j}{2} \sigma_{Z}\right]\left[U\right]^{n_j}\exp\left[-i\frac{\psi_j}{2} \sigma_{Z}\right]\right)U,
\end{equation}
where $U=\exp\left[-i \frac{5\theta_0}{2}(1+\delta)\sigma_{X}\right]$ and $\theta_0=\arccos\left[\frac{1}{4}\left(\sqrt{13}-1\right)\right]$. The error component of Eq. \eqref{eq:appendix_1} is then isolated by expanding the sequence to first order in $\delta$. The resulting unperturbed matrix and first order error matrix, $A +\delta B $,
are expressed in terms of the SU(2) generators as $A=\Lambda_1\sigma_I + i \Lambda_2\sigma_X + i \Lambda_3 \sigma_Y + i \Lambda_4\sigma_Z$ and $B=\Delta_1\sigma_I + i \Delta_2\sigma_X + i \Delta_3 \sigma_Y + i \Delta_4\sigma_Z$, where the $\Delta_i$'s and $\Lambda_i $'s are functions of $\psi_j$'s, $n_j$'s and $\theta_0$. Their closed forms are rather long to include them here, but can be easily obtained with any symbolic computation program.\\
\indent Making use again of the isomorphism between SU(2) and a subgroup of SU(4), we express the local invariants corresponding to $\mathcal{U}^{(6k)}$ in Eq. (11) in terms of the elements of the matrix $A$:
\begin{equation}
\begin{aligned}
G_1(\mathcal{U}^{(6k)})&=(\Lambda_1^2+\Lambda_4^2  - \Lambda_2^2 - \Lambda_3^2)^2\\
G_2(\mathcal{U}^{(6k)})&=3 \Lambda_4^4 + 3 \Lambda_1^4 - 2 \Lambda_1^2 (\Lambda_2^2 + \Lambda_3^2) + 3 (\Lambda_2^2 + \Lambda_3^2)^2 +
 \Lambda_4^2 (6 \Lambda_1^2 - 2 (\Lambda_2^2 + \Lambda_3^2)).
\end{aligned}
\end{equation}
\indent With the above expressions and the terms that conform the matrix $B$, we construct our objective function such that the error matrix $B$ is canceled and the local invariants of the sequence and target operation are as close as possible. Accordingly, the objective function is given by
\begin{equation}\label{eq:objective_function}
f=\Delta_1^2+ \Delta_2^2 + \Delta_3^2 + \Delta_4^2 +[G_1(\mathcal{U}^{(6k)})- G_1(\mathfrak{U})]^2 +[G_2(\mathcal{U}^{(6k)})- G_2(\mathfrak{U})]^2,
\end{equation}
where $G_i(\mathfrak{U})$ are the local invariants of the target operation.\\
\indent The values of the solutions found by numerically minimizing the objective function while targeting a {\sc cnot} operation are:
\begin{equation}
\begin{aligned}
\psi_1=& 1.135268,\\
\psi_2=& -0.405533,\\
\psi_3=& -1.841855,\\
\psi_4=& 0.191753.
\end{aligned}
\end{equation}
Moreover, the angles of the local operations needed to transform $\mathcal{U}^{(6k)}_{\text{\sc cnot}}$  into {\sc cnot}, Eq. (12), are
\begin{equation}
\begin{aligned}
\phi_1=& -1.607820,\\
\phi_2=& 0.234035.
\end{aligned}
\end{equation}
\indent Similarly, the solutions found with the numerical minimization of Eq. \eqref{eq:objective_function}  that yield a corrected rotation equivalent to $(5\theta_0/k)_{ZZ}$, for $k=\{5,10,20\}$ respectively, are
\begin{equation}
\begin{aligned}
\psi_1=\{&-0.183589,-0.103032,-0.0522225\},\\
\psi_2=\{&-3.061776,-3.129928,-3.138440\},\\
\psi_3=\{&-2.019322,-2.583841,-2.862841\},\\
\psi_4=\{&1.750803,0.844394,0.418648\}.
\end{aligned}
\end{equation}
\indent Finally, the single-qubit rotation angles in Eq. (13), for $k=\{5,10,20\}$, are
\begin{equation}
\begin{aligned}
\beta_5=&3.111045,\\
\beta_{10}=&2.290846,\\
\beta_{20}=&-1.216184,\\
\gamma_{5}=&-2.117345,\\
\gamma_{10}=&-1.850509,\\
\gamma_{20}=&1.430782.
\end{aligned}
\end{equation}

\section{II.  \  Effect of imperfect local gates on the infidelity of the composite pulse sequences}
\begin{figure}[tbp]
\centering
\begin{subfigure}[b]{0.45\linewidth}
\includegraphics[width=\linewidth]{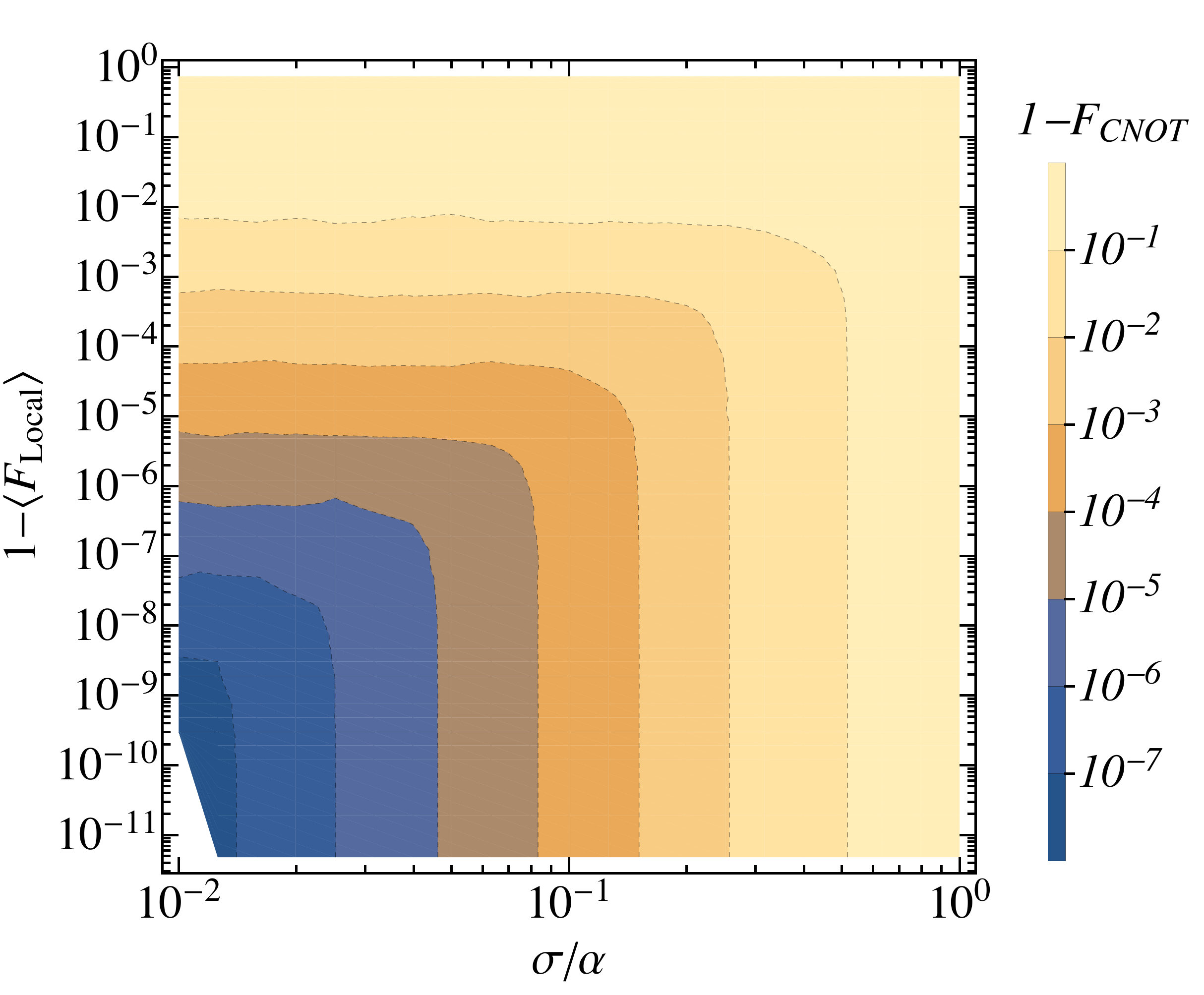}
\caption{Length-40 sequence, local systematic error }\label{fig:systematiclength40}
\end{subfigure}
\begin{subfigure}[b]{0.45\linewidth}
\includegraphics[width=\linewidth]{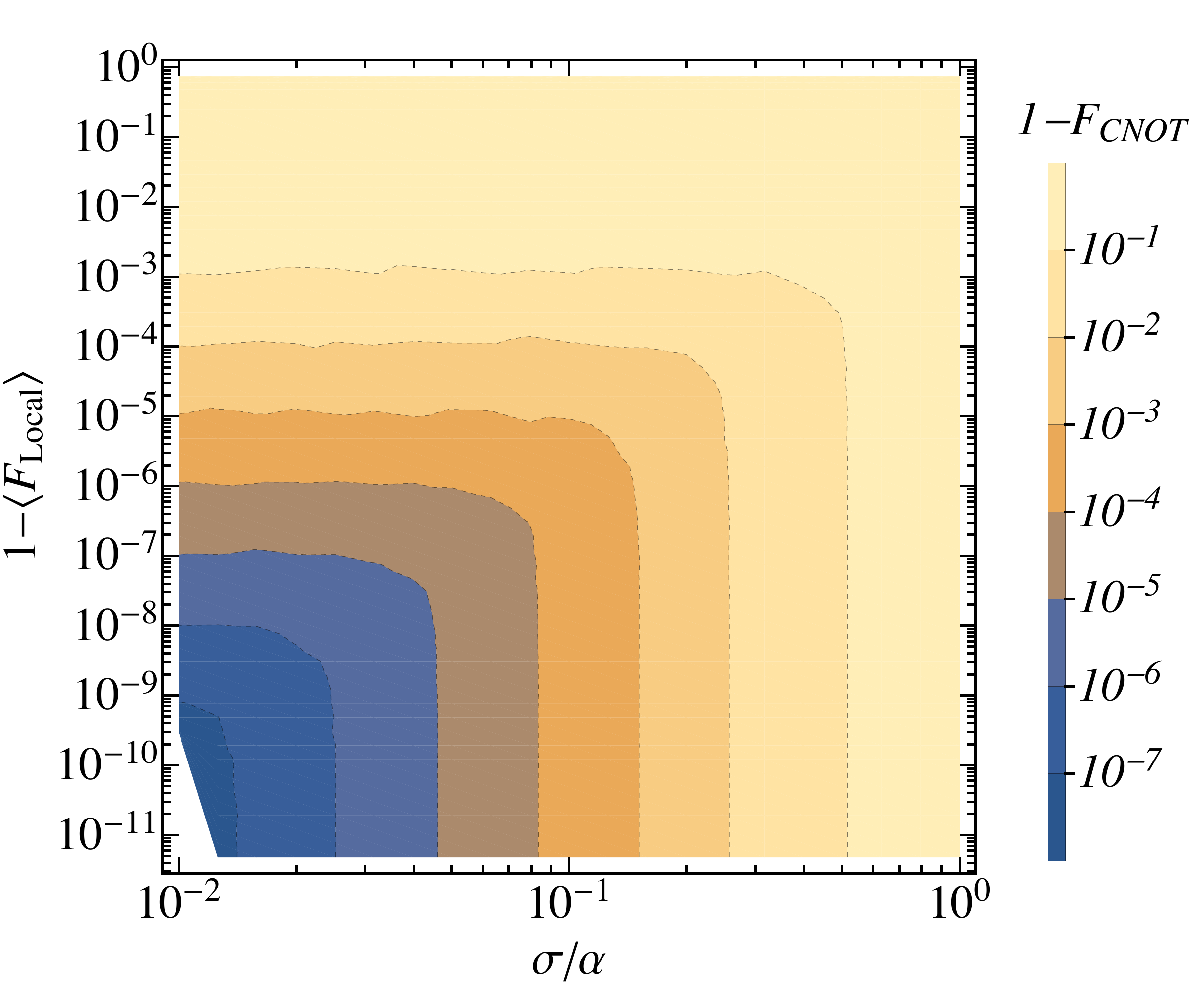}
\caption{Length-40 sequence, local random error }\label{fig:randomlength40}
\end{subfigure}
\begin{subfigure}[b]{0.45\linewidth}
\includegraphics[width=\linewidth]{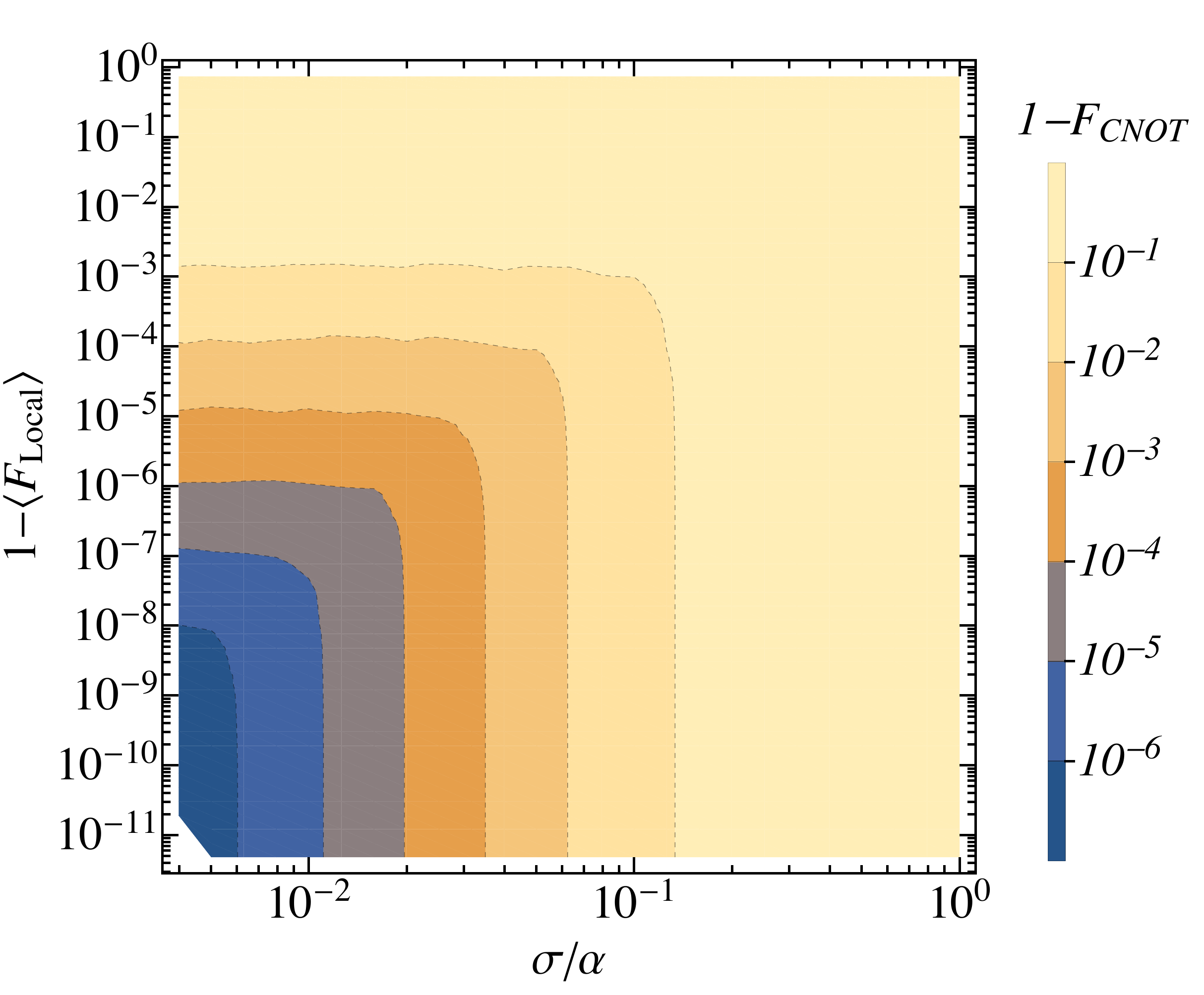}
\caption{Length-120 sequence, local systematic error }\label{fig:systematiclength120}
\end{subfigure}
\begin{subfigure}[b]{0.45\linewidth}
\includegraphics[width=\linewidth]{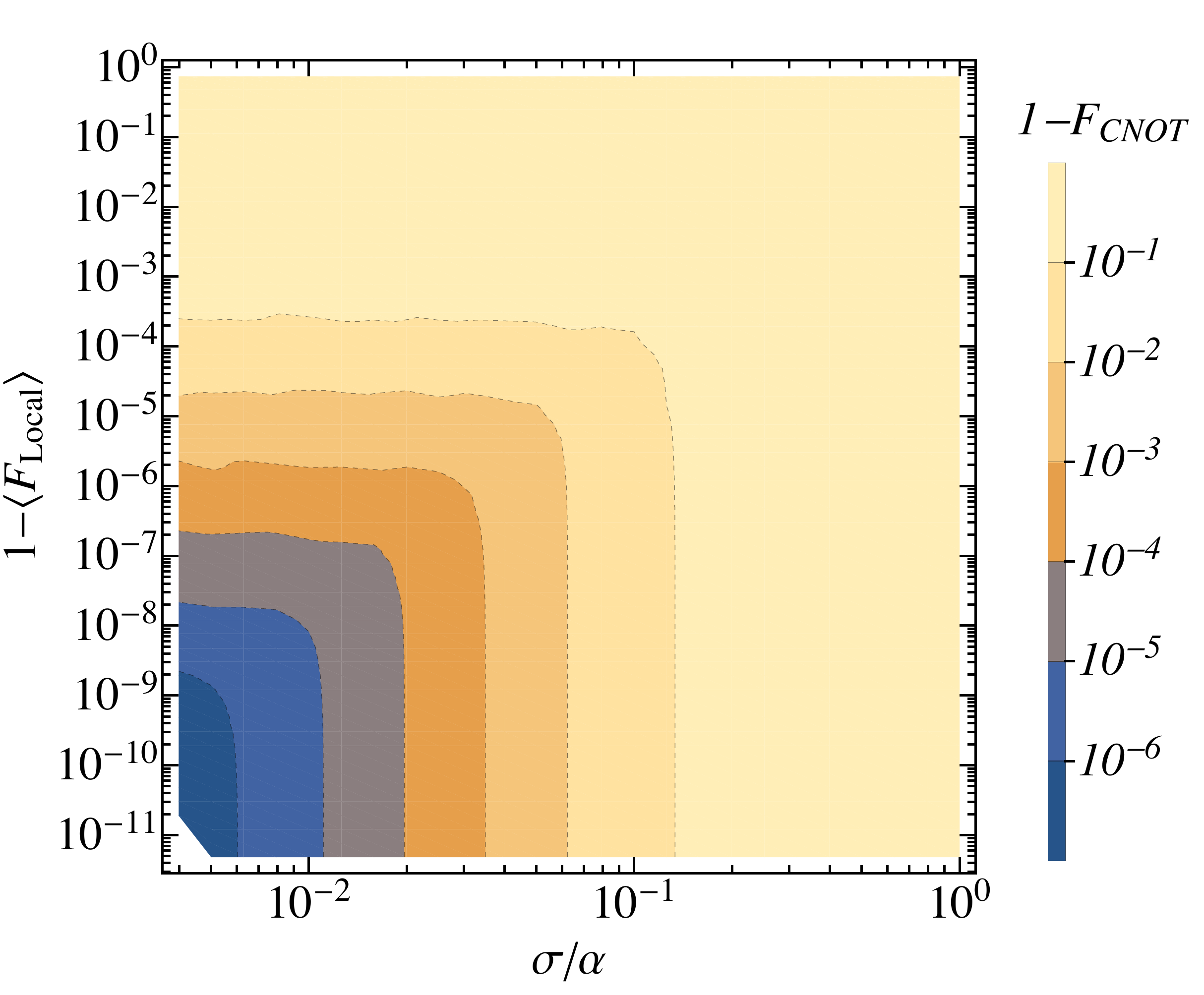}
\caption{Length-120 sequence, local random error }\label{fig:randomlength120}
\end{subfigure}
\renewcommand{\baselinestretch}{1}
\small\normalsize
\begin{quote}
\caption{(Color online) Composite {\sc cnot} infidelity vs averaged local gate infidelity vs noise strength ($\sigma/\alpha$). The length-40 sequence is given by Eq. (10) in the main text, with $k=20$. The length-120 sequence is given by Eq. (12) in the main text, with $k=20$.}
\label{fig:CNOT with CK1}
\end{quote}
\end{figure}

The contour plots in Fig. \ref{fig:CNOT with CK1} present the resulting infidelity of the length-40 (Eq. (10) with $k=20$) and length-120 (Eq. (12) with $k=20$) composite pulse sequences when systematic error in the two-qubit gate and imperfect local gates are taken into account. We apply each sequence to a Hamiltonian formed by an Ising coupling of strength $\alpha$ and random fluctuations only on the SU(4) generators that the particular sequence targets, $H= \alpha\sigma_{ZZ}+\sum \delta_{ij}\sigma_{ij}$ (the length-40 sequence targets all SU(4) generators but $\sigma_{ZZ}$, whereas the length-121 sequence targets all 15 SU(4) generators). As stated in the main text, each local gate of the composite sequence is perturbed by a random local gate of the form $\exp\left[-i \sum \Delta_i\sigma_i\right]\otimes \exp\left[-i \sum \Delta_j \sigma_j\right]$. We analyze separately two types of error that can affect the local gates: systematic and random errors. In order to represent the effect of systematic errors, when the same local gate is invoked multiple times in the sequence, it is invoked with the same perturbation. Whereas for random errors the perturbation is never the same. For each of many realizations of the perturbations, we numerically find the average infidelity of the imperfect local gates invoked as well as the infidelity of the composite pulse sequence,  which is formed using those imperfect local gates and it is also perturbed by systematic errors at the two-qubit level. These infidelities are averaged over noise realizations by sampling each stochastic noise variable over a normal distribution of standard deviation $\sigma$, with the average being taken over 500 samples for each value of $\sigma$. \\
\indent As mentioned in the main text and shown in the figures below, the average infidelity of a composite {\sc cnot} caused by errors in the local gates increases with the length of the sequence. For systematic local errors, the {\sc cnot} infidelity increases up to about 80 times the local gate infidelity for the length-120 sequence, which has 121 local gates. On the other hand, random local errors have qualitatively the same effect as systematic ones, but are quantitatively more pernicious, resulting in a {\sc cnot} infidelity increase up to about 480 times the random local gate infidelity for the length-120. Similarly, the {\sc cnot} formed using the length-40 sequence has 41 local gates and about 18 times the local gate infidelity for systematic errors and about 90 times the local gate infidelity for random errors. Fortunately, random errors are typically much smaller to begin with than systematic errors.

\section{III. \ Improving the gate fidelity of the cross resonance gate between transmon qubits}
In a recent experimental work with the cross resonance (CR) gate between transmon qubits by Sheldon et al. \cite{Sheldon2016}, the two-qubit entangling gate was improved through a detailed Hamiltonian estimation and the application of a cancellation tone that raises the two-qubit fidelity over 99$\%$. The experimental CR Hamiltonian is stated in terms of $\sigma_{ZX}, \ \sigma_{ZY}, \ \sigma_{ZZ}, \ \sigma_{IX}, \ \sigma_{IZ}$, and $ \sigma_{IY}$, of which $\sigma_{ZZ}$ is comparatively small and $\sigma_{IZ}$ is negligible. The authors improve the CR gate by applying a cancellation tone on the target qubit such that unwanted interaction terms of the CR Hamiltonian are eliminated. They choose the cancellation tone phase at which $\sigma_{ZY}$ and $\sigma_{IY}$ are zero, and the amplitude of the cancellation tone is tuned such that $\sigma_{IX}$ and $\sigma_{IY}$ are zero as well. With this technique, the authors report a $\sigma_{ZX}$ gate that is locally equivalent to {\sc cnot} with gate fidelity of 99.1$\%$, an important improvement from previously reported fidelities of 94-96$\%$.\\
\indent Nonetheless, as stated in their work, there are systematic error terms that still remain after the experimental procedure. According to their modeling, the residual error corresponds to a $\sigma_{ZZ}$ term  in the Hamiltonian around an order of magnitude smaller than the interaction term, and a $\sigma_{IX}$ term an order of magnitude smaller than $\sigma_{ZZ}$. Following the method presented in the main text, we transform this Hamiltonian into a ZZ coupling by applying a Hadamard transformation to the second qubit. In this context, the dominant error term will give two error channels from the anticommuting set, which can be corrected with the length-5 sequence, Eq. (10) with $k=5$. Using the experimentally reported parameters, and considering that with the length-5 sequence the {\sc cnot} infidelity is about 8 times the average single-qubit gate infidelity, we calculate that our sequence would immediately improve the {\sc cnot} fidelity from the current value of 99.1$\%$ up to 99.6$\%$ with the presently achievable single-qubit gate fidelities of 99.95$\%$ \cite{Sheldon2016a}. All this was calculated with the $\sigma_{IX}$ error present, which, in principle, can be completely corrected from the start by a more precisely tuned cancellation tone. If one were to improve the single-qubit gate fidelities and $T_2$ time, our sequence could further boost the two-qubit fidelity up to 99.98$\%$. 

\renewcommand{\baselinestretch}{2}
\small\normalsize

\putbib[library]

\end{bibunit}

\end{document}